\documentclass[letterpaper]{article} 
\usepackage{aaai2027}  
\usepackage[hyphens]{url}  
\usepackage{graphicx} 
\usepackage{natbib} 
\usepackage{caption} 
\usepackage{booktabs}
\usepackage{multirow}

\title{IteraSong: LLM-Guided Iterative Symbolic Music Creation in a Web DAW}

\author{
    Di Lu\textsuperscript{\rm 1,2,3},
    Xinran Zhang\textsuperscript{\rm 1,2,3},
    Yan Gao\textsuperscript{\rm 1,2,3},
    Yuan Zhang\textsuperscript{\rm 1,2,3},
    Xiaoqing Wang\textsuperscript{\rm 1,2,3},
    Qingwen Zhou\textsuperscript{\rm 1,2,3},
    Xiaobing Li\textsuperscript{\rm 1,2,3},
    Maosong Sun\textsuperscript{\rm 1,2,3,4}\corresponding
}
\affiliations{
    \textsuperscript{\rm 1}Department of AI Music and Music Information Technology, Central Conservatory of Music\\
    \textsuperscript{\rm 2}Key Laboratory of Music and Brain Science, Central Conservatory of Music, Ministry of Education, China\\
    \textsuperscript{\rm 3}Laboratory of AI Music, Central Conservatory of Music, Laboratory of Philosophy and Social Sciences, Ministry of Education, China\\
    \textsuperscript{\rm 4}Department of Computer Science and Technology, Tsinghua University\\
    ld@ccom.edu.cn, sms@tsinghua.edu.cn
}

\begin{document}

\maketitle

\begin{abstract}
Music creation is inherently iterative, yet many symbolic generation systems still center on producing or regenerating isolated outputs. 
We present \textbf{IteraSong}, an interactive web-based Digital Audio Workstation (DAW) for LLM-guided iterative symbolic music creation.
Unlike one-shot generation systems, IteraSong supports a generate–audition–edit–refine loop over a persistent multi-track project.
Users initialize a fully labeled pop draft, then progressively refine it through natural-language instructions.
The LLM maps requests to a hierarchical action space of editing primitives exposed via Model Context Protocol (MCP) tools, spanning global, track, section, and bar/note scopes.
The demo enables attendees to edit multi-track compositions in natural language, apply deterministic and generative edits, and undo/redo changes in real time.
\end{abstract}

\begin{links}
     \link{Video}{https://youtu.be/XZaHHms5Kl0}
\end{links}

\section{Introduction}

Symbolic music generation has advanced rapidly, enabling controllable generation of multitrack compositions with explicit musical structure~\cite{vonrutte2023figaro,pasquier2025midigpt}. 
Recent LLM-based approaches further show that language models can understand symbolic music and support music-related reasoning and generation~\cite{yuan2024chatmusician,yu2023musicagent}. 
However, practical music creation is rarely a one-shot process: creators typically listen, revise, compare, and refine a draft over multiple rounds. 
This motivates interactive tools that support iterative creation over an evolving musical project rather than treating generation as the final step.

Existing interactive symbolic music systems mainly focus on localized infilling or structured composition~\cite{malandro2023composer,malandro2024composer2,hu2025compose,he2025tomi}, while LLM-based music systems explore natural-language interaction and tool orchestration~\cite{yuan2024chatmusician,yu2023musicagent}. 
Yet iterative creation requires more than isolated generation or tool calls: each edit should build on the current project, modify only the intended musical content, and allow unsatisfactory changes to be revised or undone. 
Few existing systems integrate these capabilities into a unified project-level workflow for multi-round natural-language symbolic music editing.

We present \textbf{IteraSong}, a Web DAW for iterative symbolic music creation.
Our main contributions are:

\begin{itemize}

\item A project-centered interactive system that enables multi-round, project-level symbolic music editing via natural language.

\item A hierarchical action space of editing primitives exposed through MCP-compatible tools, enabling precise, scoped edits from global structure down to individual notes.

\item An end-to-end workflow integrating initial draft generation, LLM-guided editing, and in-browser playback, designed for live hands-on interaction.

\end{itemize}

\section{System Design}
\begin{figure*}[t]
\centering
\includegraphics[width=0.95\textwidth]{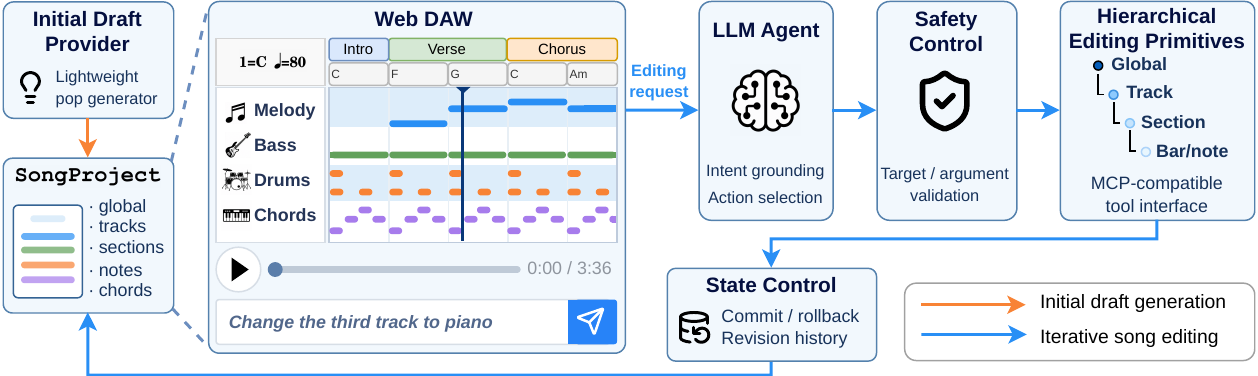}
\caption{Overall IteraSong architecture for initial draft generation and iterative song editing in the Web DAW.}
\label{fig:architecture}
\end{figure*}

Figure~\ref{fig:architecture} illustrates the IteraSong workflow. 
First, a lightweight pop generator produces a fully labeled multitrack draft, which is normalized into a \texttt{SongProject} and maintained as the persistent project state throughout iterative creation. 

During each subsequent interaction round, the user submits a natural language editing request. The LLM interprets the request, selects one or more editing primitives, and invokes them through MCP-compatible tools~\cite{hou2026mcp}.

After successful execution, the updated project is immediately available for audition in the Web DAW. The user may continue editing, request another variation, or undo a previous change. Each accepted edit becomes the starting state for the next interaction round. 

\paragraph{Initial Draft Provider.}
IteraSong starts from an editable symbolic draft rather than an empty project. 
The current prototype uses a lightweight pop generator as a controllable testbed, while the framework can also accept drafts from more advanced symbolic generation models.
The generator provides global attributes (e.g. key, mode, tempo), track identities and roles, section boundaries, individual note events, and chord progressions. 
The resulting \texttt{SongProject} therefore exposes explicit musical objects that can be visualized in the Web DAW and addressed during subsequent editing.

\paragraph{LLM Agent.}
Rather than directly manipulating raw symbolic song data, the LLM serves as an intent interpreter and action selector. Given a natural-language request, it identifies the intended musical target and operation, then invokes the corresponding MCP-compatible tool.
For example, ``change the third track to piano'' maps to the deterministic primitive \texttt{set\_instrument}, whereas ``regenerate the bass in Chorus~2'' maps to the section-scoped generative primitive \texttt{regenerate\_section}.

\paragraph{Safety Control.}
Before any mutation, the safety module checks whether the requested targets exist and are unambiguous, whether the arguments are valid, and whether the requested operation is supported. Requests that fail these checks are rejected without modifying the project.

\paragraph{Hierarchical Editing Primitives.}
IteraSong exposes a bounded set of music-domain primitives organized at global, track, section, and bar/note levels. 
The action space includes observation and DAW-control primitives,
together with deterministic and scoped generative editing primitives.
Observation primitives expose project, track, structure, chord, and note information, while DAW controls provide play, stop, seek, loop, mute, and solo operations. Deterministic and generative primitives are summarized in Table~\ref{tab:primitives}.

\begin{table}[th]
\centering
\small
\setlength{\tabcolsep}{3pt}
\begin{tabular}{@{}lp{0.42\columnwidth}p{0.37\columnwidth}@{}}
\toprule
\textbf{Level} & \textbf{Deterministic} & \textbf{Generative} \\
\midrule
Global
& set\_tempo, set\_key\_mode, song\_transpose
& change\_entry\_strategy, change\_ending\_strategy \\

Track
& set\_instrument, add\_track, clear\_track
& regenerate\_track \\

Section
& set\_chords, clear\_section, transpose\_section
& regenerate\_section, reharmonize\_section \\

Bar/note
& add\_note, delete\_note, transpose\_note
& regenerate\_note \\
\bottomrule
\end{tabular}
\caption{Hierarchical editing primitives.}
\label{tab:primitives}
\end{table}

\paragraph{State Control.}
State control manages the evolving project across interaction rounds.
Each symbolic mutation is applied transactionally to the current project revision: successful edits create a new revision and history entry, whereas failed operations leave the previous state unchanged. 
This mechanism also supports undo and rollback, allowing later interaction rounds to continue from a valid and explicitly selected project state.

\section{Demonstration Scenario}

The demonstration presents a continuous creation session rather than isolated tool calls. 
A user first generates a labeled multitrack pop song with sections, chords, and instrument tracks in one click.
After auditioning the draft, the user can issue successive natural-language requests to change an instrument, regenerate the accompaniment in Chorus~2, request another variation, undo/redo edits, or correct an individual note.
Each accepted edit is visualized and auditioned, while the same project is retained for subsequent requests. 
Users can also use revision history to revisit previous versions.

\section{Implementation and Validation}

\paragraph{Implementation.}
IteraSong uses a rule-based pop generator for both initial draft generation and generative editing. 
The Web DAW uses a Node.js backend for symbolic visualization,
note-level interaction, and natural-language chat.
Browser-side MIDI playback uses FluidSynth compiled to WebAssembly (WASM).
Qwen-Max serves as the LLM agent, invoking observation, DAW-control,
deterministic, and generative primitives through MCP-compatible tools.

\paragraph{Validation.}
We evaluated 20 ten-turn sessions (200 turns) across five \texttt{SongProject}s and four workflows, covering Chinese and English requests. Using Qwen-Max and the actual MCP execution path,
IteraSong achieved 90.63\% task completion (145/160), 100\% safe rejection (40/40), and 100\% state consistency (200/200).

\section{Conclusion}

IteraSong enables iterative symbolic music creation through structured project state, hierarchical editing primitives, and natural-language interaction in a Web DAW.
The current prototype focuses on pop music with a rule-based generation backend.
Perceptual quality and user experience remain for future evaluation.
Future work will also extend the action space and integrate stronger generative backends.

\section*{Acknowledgments}
This work was supported by the Special Program of the National Natural Science Foundation of China (Grant No. T2341003), the Advanced Discipline Construction Project of Beijing Universities, and the Central Conservatory of Music General Research Project (Grant No.25YB14). We thank our colleagues for their helpful discussions and feedback on this work.

\bibliography{iterasong}

\clearpage
\appendix
\setcounter{secnumdepth}{2}

\section*{Appendix}

\section{Multi-round Evaluation Details}

\paragraph{Test projects and session design.}
The evaluation used 5 symbolic pop projects from a shared generation pipeline, spanning 11--12 tracks, 14--16 sections, and 6,241--8,536 note events. Each project was tested with 4 ten-turn workflows: \emph{track-first}, \emph{section-first}, \emph{note-first}, and \emph{mixed editing}. 
The resulting 20 sessions comprised 10 in Chinese and 10 in English. 
Each session contained 1 observation, 7 mutation requests (including editing and undo/redo), and 2 invalid requests.
Qwen-Max executed the requests through the stdio MCP client--server path. 
Project state was preserved throughout each session, without selective retries or resets.

\paragraph{Per-session outcomes.}
Table~\ref{tab:multiround_sessions} expands the aggregate validation results in the main paper by showing the completion count for every project--workflow combination. All observation requests (20/20) and invalid-request rejections (40/40) succeeded, while unsuccessful requests were concentrated in 3 sessions.

\begin{table}[th]
\centering
\small
\begin{tabular}{@{}lccrrr@{}}
\toprule
& & & \multicolumn{3}{c}{Success / Total} \\
\cmidrule(l){4-6}
Project & Workflow & Turns
& Obs. & Mut. & Rej. \\
\midrule
\multirow{4}{*}{Project 1}
 & 1 & 10 & 1/1  & 7/7   & 2/2 \\
 &  2 & 10 & 1/1  & 7/7   & 2/2 \\
 &  3 & 10 & 1/1  & 7/7   & 2/2 \\
 &  4 & 10 & 1/1 & 7/7   & 2/2 \\
\midrule
\multirow{4}{*}{Project 2}
 &  1 & 10 & 1/1 & \textbf{1/7} & 2/2 \\
 &  2 & 10 & 1/1 & 7/7   & 2/2 \\
 &  3 & 10 & 1/1 & 7/7   & 2/2 \\
 &  4 & 10 & 1/1 & 7/7   & 2/2 \\
\midrule
\multirow{4}{*}{Project 3}
 &  1 & 10 & 1/1 & 7/7   & 2/2 \\
 &  2 & 10 & 1/1 & \textbf{2/7} & 2/2 \\
 &  3 & 10 & 1/1 & 7/7   & 2/2 \\
 &  4 & 10 & 1/1 & \textbf{3/7} & 2/2 \\
\midrule
\multirow{4}{*}{Project 4}
 &  1 & 10 & 1/1 & 7/7   & 2/2 \\
 &  2 & 10 & 1/1 & 7/7   & 2/2 \\
 &  3 & 10 & 1/1 & 7/7   & 2/2 \\
 &  4 & 10 & 1/1 & 7/7   & 2/2 \\
\midrule
\multirow{4}{*}{Project 5}
 &  1 & 10 & 1/1 & 7/7   & 2/2 \\
 &  2 & 10 & 1/1 & 7/7   & 2/2 \\
 &  3 & 10 & 1/1 & 7/7   & 2/2 \\
 &  4 & 10 & 1/1 & 7/7   & 2/2 \\
\midrule
\multicolumn{2}{l}{\textbf{Total (20 sessions)}} & \textbf{200}
 & \textbf{20/20} & \textbf{125/140} & \textbf{40/40} \\
\bottomrule
\end{tabular}
\caption{Per-session results for the multi-round editing evaluation.
Each row is 1 ten-turn session on a single \texttt{SongProject}. 
Workflows 1--4 correspond to track-first, section-first, note-first, and mixed editing, respectively.
Obs., Mut., and Rej. denote observation requests, mutation requests, and invalid-request rejections, respectively.}
\label{tab:multiround_sessions}
\end{table}

\paragraph{Representative ten-turn session.}
Table~\ref{tab:appendix_session} summarizes one recorded English section-first session. It illustrates successive edits to the same project, revision changes after undo, and unchanged revisions following rejected requests. Requests are shortened for readability.

\begin{table}[t]
\centering
\footnotesize
\setlength{\tabcolsep}{3pt}
\begin{tabular}{@{}c p{0.50\columnwidth} p{0.30\columnwidth} c@{}}
\toprule
\textbf{Turn} & \textbf{Request} & \textbf{Observed effect} & \textbf{Rev.} \\
\midrule
1 & Inspect the structure(read-only). & - & 1\\
\midrule
2 & Raise chorus~2 and its chords by two semitones & 316 notes and chords changed. & 2\\
\midrule
3 & Undo & Restored the original notes and harmony. & 3\\
\midrule
4 & Regenerate track~3 in chorus~2 & 108 replacement notes with rhythm changed. & 4\\
\midrule
5 & On track~2, raise the pitch-33 note at bar~2, beat 1 to 34  & 1 note changed. & 5\\
\midrule
6 & Raise chorus 99 by two semitones & Rejected without mutation. & 5\\
\midrule
7 & Raise track~2 by 127 semitones & Rejected without mutation. & 5\\
\midrule
8 & Set tempo to 100 BPM & Committed. & 6\\
\midrule
9 & Raise track~2 by one semitone & 811 notes transposed. & 7\\
\midrule
10 & Undo the track transposition & Restored the prior musical content. & 8\\
\bottomrule
\end{tabular}
\caption{One recorded section-first session. Revisions are shown after each turn.}
\label{tab:appendix_session}
\end{table}

\paragraph{Scoring and failure analysis.}
Scoring required each turn to reach its intended state without modifying content outside the authorized scope.
Dependent turns additionally had to satisfy their predecessor conditions. 
The 15 incomplete mutation turns consisted of 3 primary target-binding errors and 12 downstream dependency failures.
In 3 Chinese-language regeneration requests, the agent supplied \texttt{section\_8} instead of the correctly resolved \texttt{section\_9}, thus validation rejected these mismatches before mutation.

\section{Web DAW Interface}
Figure~\ref{fig:ui} illustrates how the evolving \texttt{SongProject} is presented during editing. 
Tracks, section boundaries, chords, and notes remain visible alongside the agent's requests and execution trace, allowing the user to inspect project changes while auditioning the current arrangement.

\section{Scope and Limitations}
This functional evaluation was conducted on fixed development workflows.
Since the projects share a common generation backend and turns within each session are dependent, the results are descriptive and do not establish generalization..
The evaluated build used a scoped regeneration adapter.
Furthermore, the evaluation did not assess audible playback,
perceived music quality, or usability with professional musicians.

\twocolumn[{%
\centering
\includegraphics[width=\textwidth]{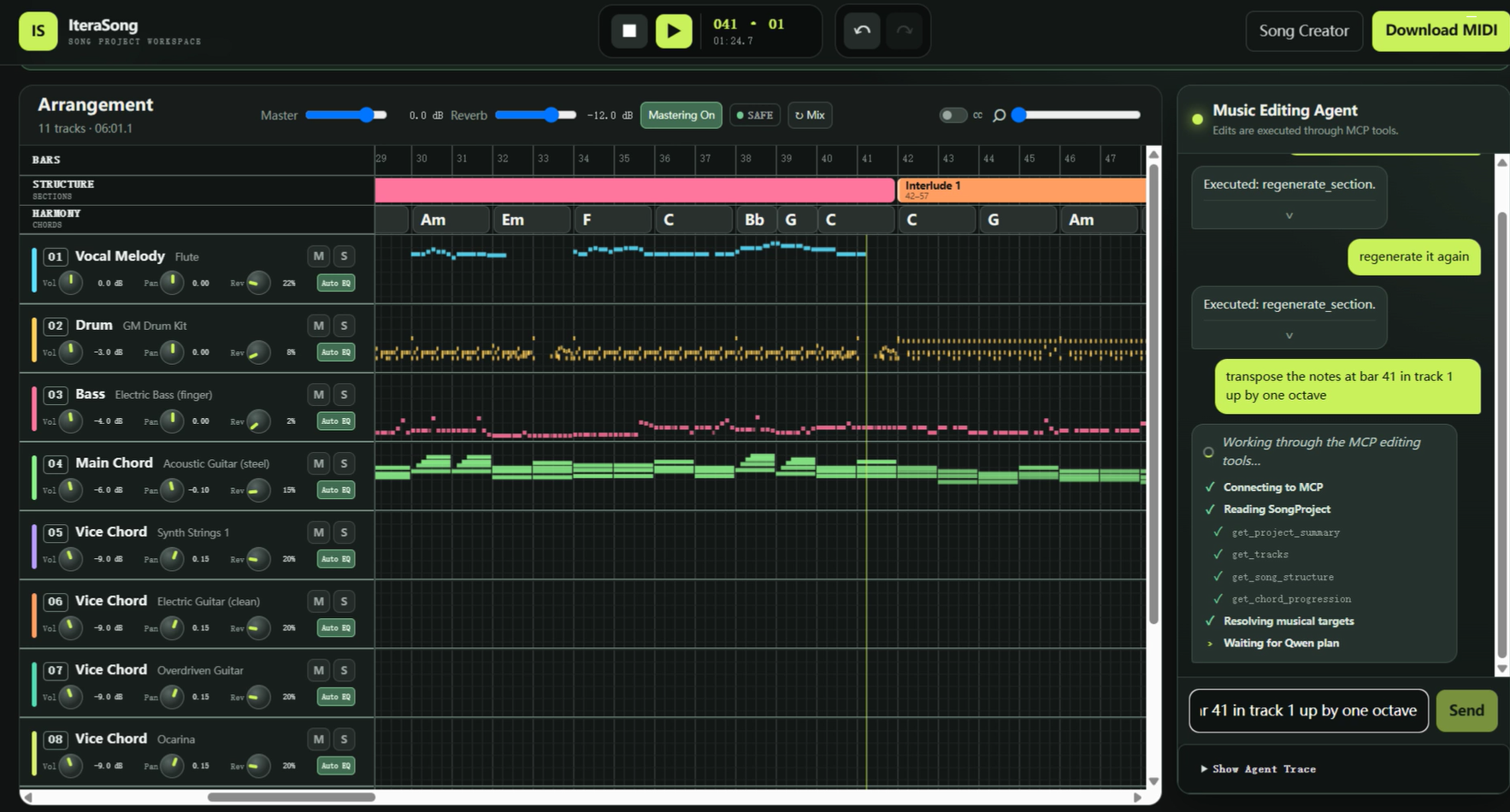}\par
\captionof{figure}{IteraSong Web DAW during a multi-round editing session. The arrangement window displays symbolic project content, while the agent panel shows natural-language requests and MCP tool execution.}
\label{fig:ui}
\vspace{0.8em}
}]

\end{document}